\documentclass{PoS}
\usepackage{amsmath,amssymb,amsthm,bm,bbm,color}
\usepackage{graphicx,psfrag,subfigure,rotating}

\title{Chiral Symmetry Breaking from Center Vortices}

\ShortTitle{Chiral Symmetry Breaking from Center Vortices}

\author{\speaker{Roman H\"ollwieser}\\Institute of Atomic and Subatomic Physics, Vienna University of Technology, Austria and\\
Department of Physics, New Mexico State University, Las Cruces, NM 88003-8001, USA\\E-mail: \email{hroman@kph.tuwien.ac.at}}
\author{Manfried Faber\\Institute of Atomic and Subatomic Physics, Vienna University of Technology, Austria\\E-mail: \email{faber@kph.tuwien.ac.at}}
\author{Thomas Schweigler\\Institute of Atomic and Subatomic Physics, Vienna University of Technology, Austria}

\author{Urs M. Heller\\American Physical Society, One Research Road, Ridge, NY
11961, USA\\E-mail: \email{heller@aps.org}}

\abstract{We analyze the creation of near-zero modes from would-be zero modes of
various topological charge contributions from classical center vortices in
$SU(2)$ lattice gauge theory. We show that colorful spherical vortex and
instanton configurations have very similar Dirac eigenmodes and also vortex
intersections are able to give rise to a finite density of near-zero modes,
leading to chiral symmetry breaking via the Banks-Casher formula. We discuss the
influence of the magnetic vortex fluxes on quarks and how center vortices may
break chiral symmetry.\footnote{This research was partially supported by the
Austrian Science Fund (FWF) under contract P22270-N16 (R.H.).}~~\footnote{The
numerical calculations were performed at the Vienna Scientific Cluster (VSC).}}

\FullConference{31st International Symposium on Lattice Field Theory - LATTICE 2013\\July 29 - August 3, 2013\\Mainz, Germany}

\begin{document} 

\section{Introduction}
A well established theory of spontaneous chiral symmetry breaking ($\chi$SB) relies on instantons, which are localized 
in space-time and carry a topological charge of modulus $1$. According to the 
Atiyah-Singer index theorem, a zero mode of the Dirac operator arises, which is 
concentrated at the instanton core. In the instanton liquid model overlapping 
would-be zero modes split into low-lying nonzero modes which create the chiral
condensate. Center vortices are promising candidates for explaining confinement. 
The vortex model of confinement is theoretically appealing and was confirmed by a 
multitude of numerical calculations, both in lattice Yang-Mills theory and 
within a corresponding infrared effective model, see {\it e.g.}~\cite
{DelDebbio:1996mh,Engelhardt:1999wr}. Lattice simulations indicate that vortices 
are responsible for topological charge and $\chi$SB as well~\cite
{deForcrand:1999ms,Engelhardt:2002qs,Hollwieser:2008tq}, and thus unify all 
nonperturbative phenomena in a common framework. A similar picture to the 
instanton liquid model exists insofar as lumps of topological charge arise at the 
intersection and writhing points of vortices.  The colorful, spherical SU(2) 
vortex was introduced in a previous article of our group~\cite{Jordan:2007ff} and 
may act as a prototype for this picture, as it contributes to the topological 
charge by its color structure, attracting a zero mode like an instanton. We show 
how the interplay of various topological structures from center vortices (and 
instantons) leads to near-zero modes, which by the Banks-Casher relation are 
responsible for a finite chiral condensate. 

\section{Free Dirac eigenmodes}
\label{sec:modes}

\begin{figure}[hb]
	\centering
		\includegraphics[width=.32\columnwidth]{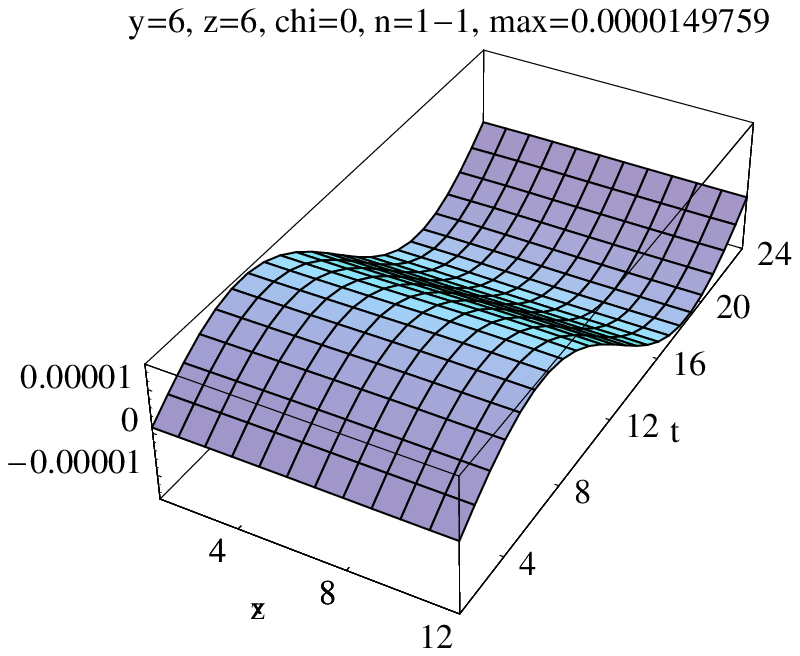}
		\includegraphics[width=.32\columnwidth]{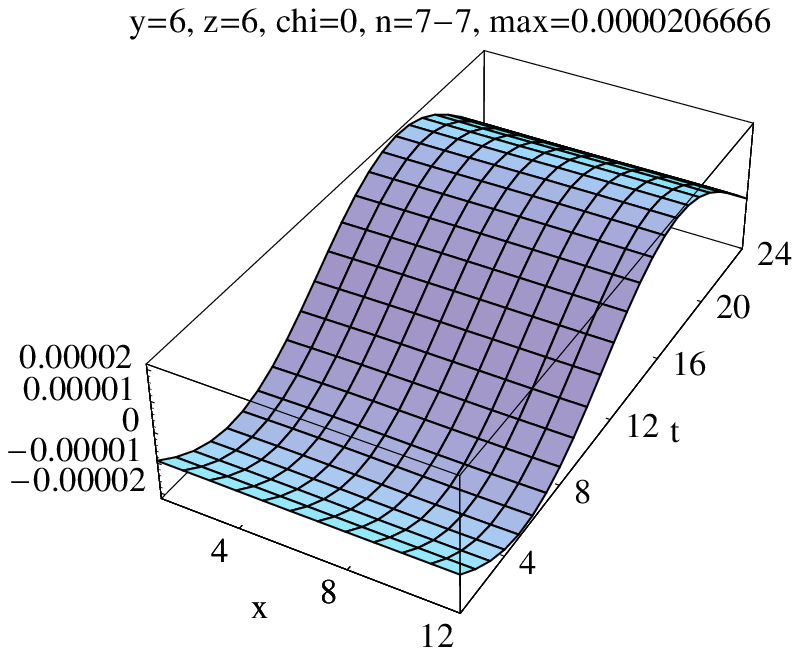}
		\includegraphics[width=.32\columnwidth]{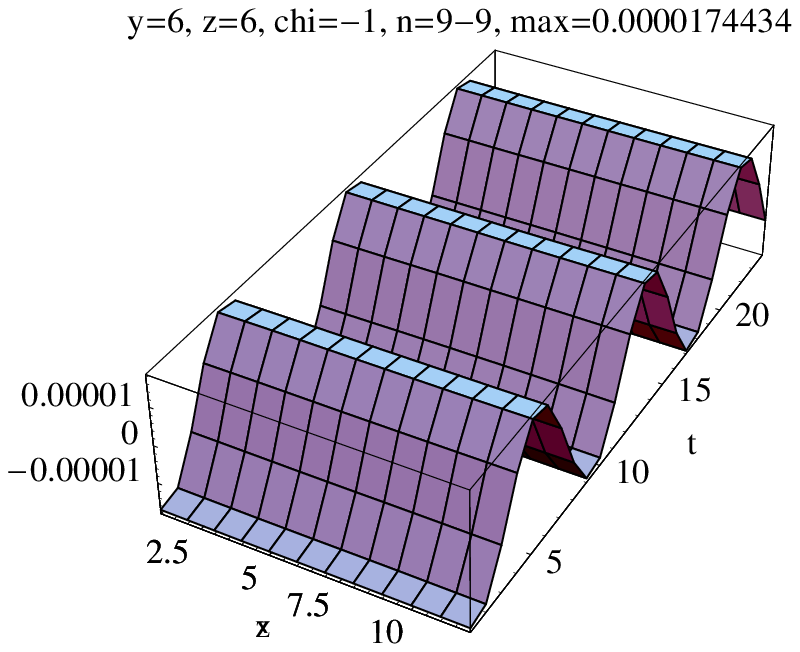}
	\caption{Chiral density of the low-lying eigenmodes of the free
	overlap Dirac operator: $\rho_5\#1$ (left), $\rho_5\#7$ (center)
	$\rho_5\#9$ (right). The modes clearly show the plane wave
behavior with oscillations of $2p_\mu$ (see text).}
	\label{fig:freeoverlap}
\end{figure}

The chiral density of free overlap eigenmodes obtained numerically using the MILC 
code are shown in Fig.~\ref{fig:freeoverlap}. The modes are found with the Ritz 
functional algorithm with random start and for degenerate eigenvalues the 
eigenmodes span a randomly oriented basis in the degenerate subspace.
Therefore the numerical modes presented in Fig.~\ref{fig:freeoverlap} are 
linear combinations of plane waves with momenta $\pm p_\mu$ and show plane wave 
oscillations of $2p_\mu$ in the chiral density. For $12^3\times24$ lattices, the first eight degenerate modes consist of plane waves with $p_4=\pm\pi/24$, hence there is one sine (cosine) oscillation in time direction, the next eight have $p_4=\pm3\pi/24$, 
{\it i.e.}, three oscillations in the time direction. The oscillations of $
\chi_R$ and $\chi_L$ are separated by half an oscillation length, {\it i.e.}, 
the maxima of $\rho_+$ correspond to minima of $\rho_-$ and vice versa,
accordingly, the scalar density is constant as expected for free eigenmodes.

\section{The Colorful Spherical Vortex}\label{sec:spher}

\begin{figure}[hb]
	\centering
		a)\includegraphics[width=.47\columnwidth]{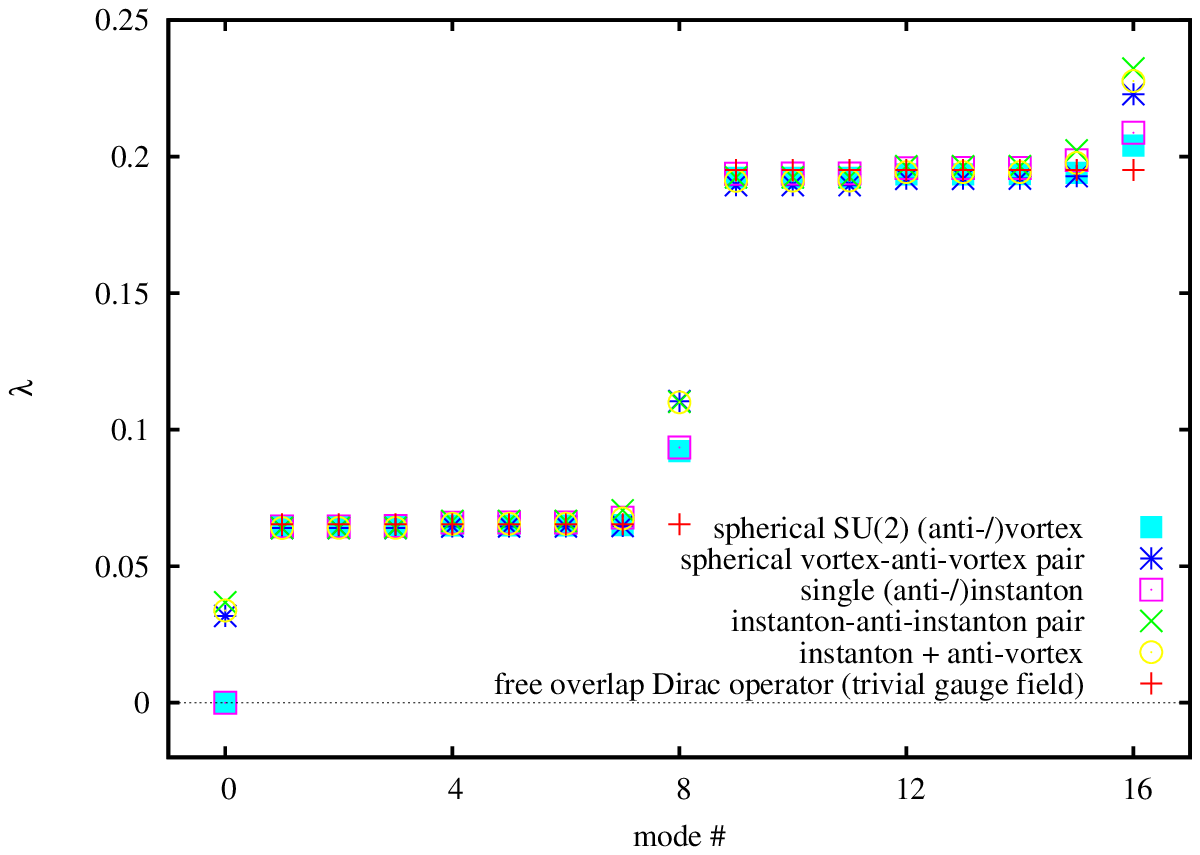}
		b)\includegraphics[width=.47\columnwidth]{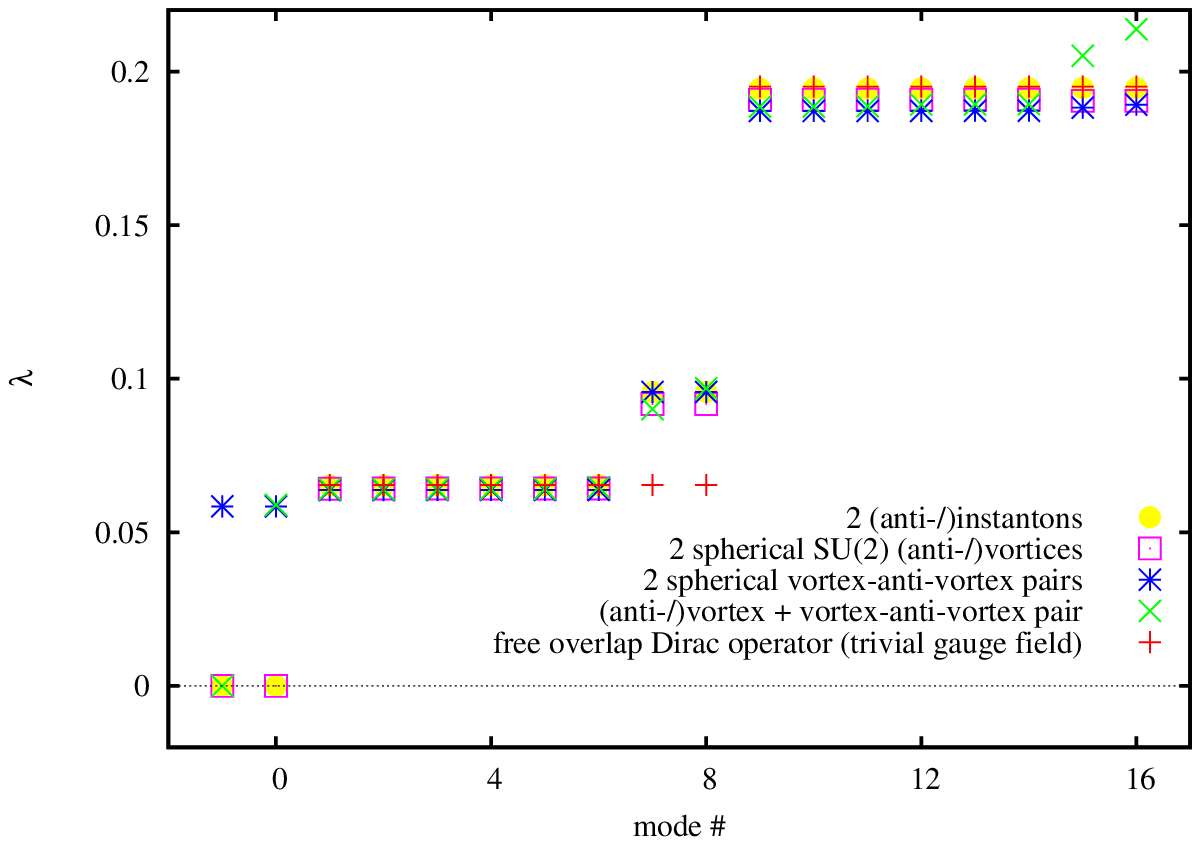}
	\caption{The lowest overlap eigenvalues for instanton and spherical
	vortex configurations compared to the eigenvalues of the free
	(overlap) Dirac operator.}
	\label{fig:oevlsph}
\end{figure}

\begin{figure}
	\centering
		a)\includegraphics[width=.32\columnwidth]{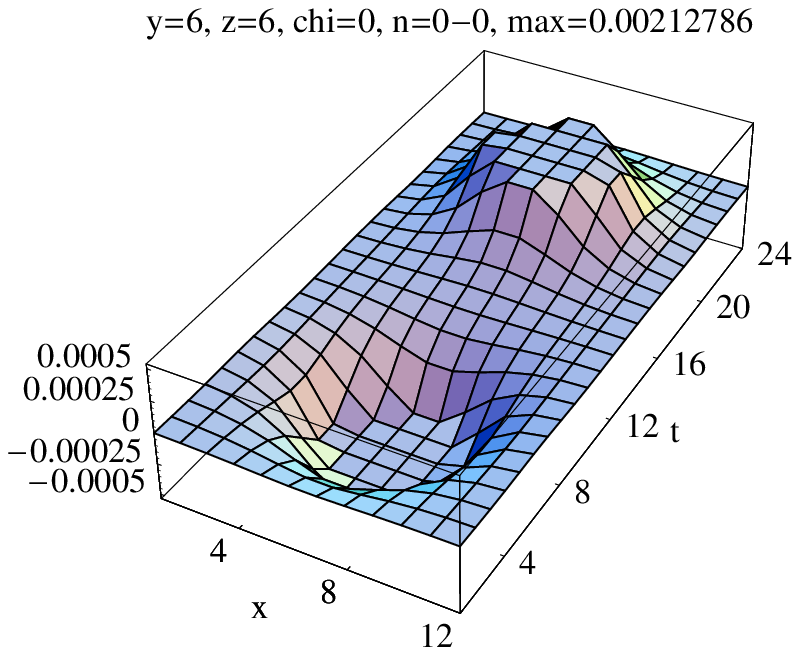}
		\includegraphics[width=.32\columnwidth]{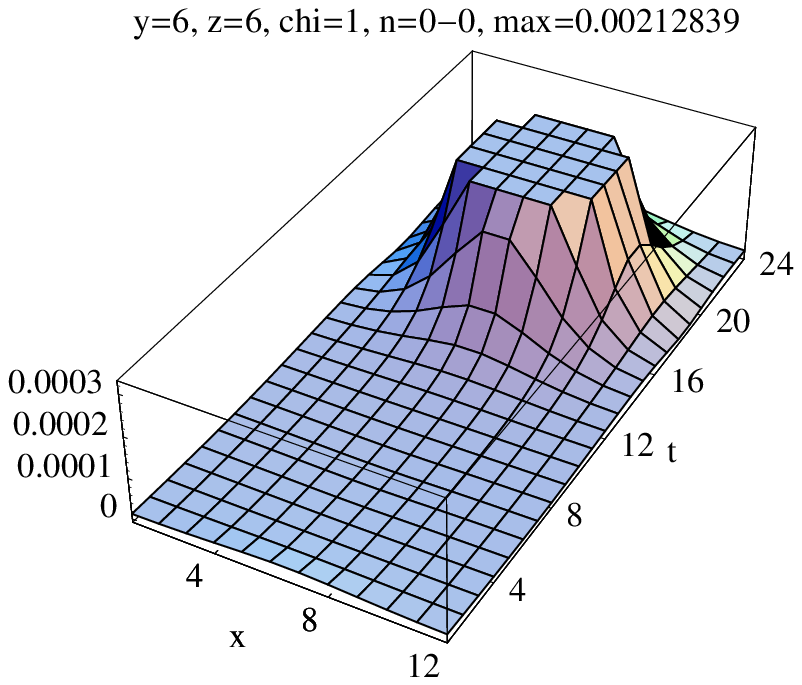}
		\includegraphics[width=.32\columnwidth]{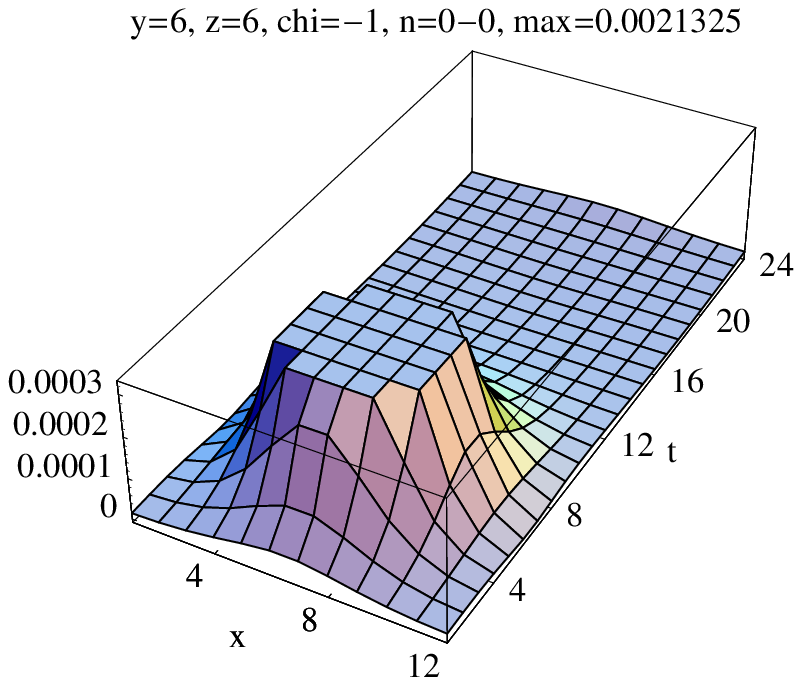}\\
		b)\includegraphics[width=.32\columnwidth]{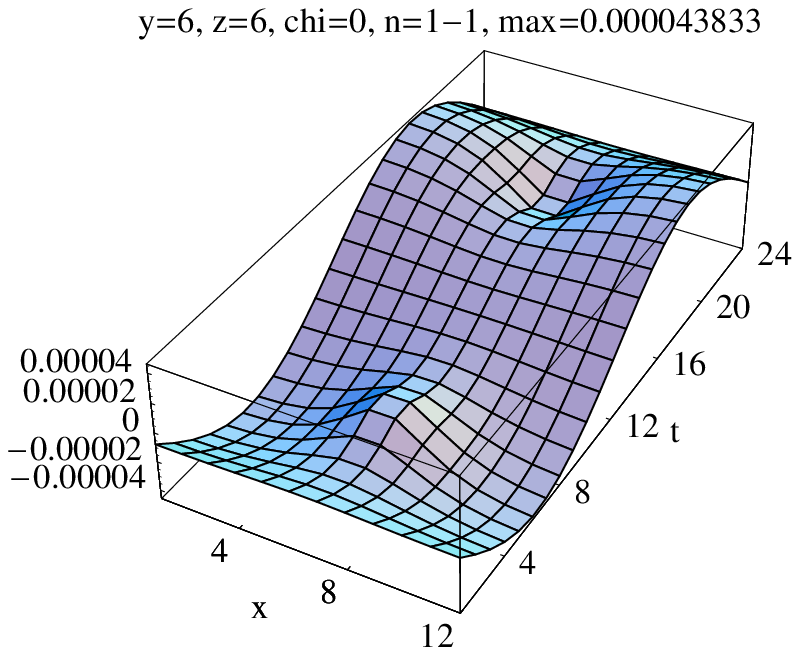}
		\includegraphics[width=.32\columnwidth]{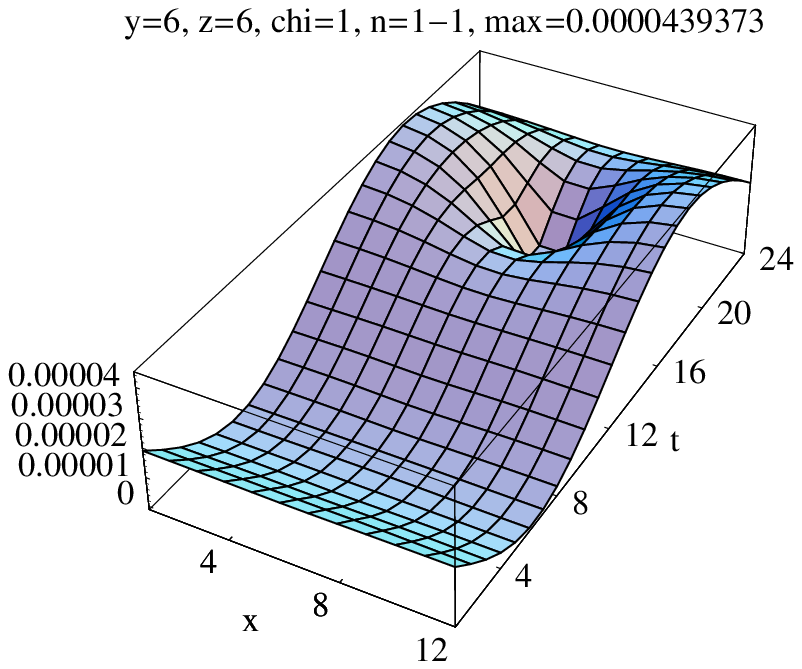}
		\includegraphics[width=.32\columnwidth]{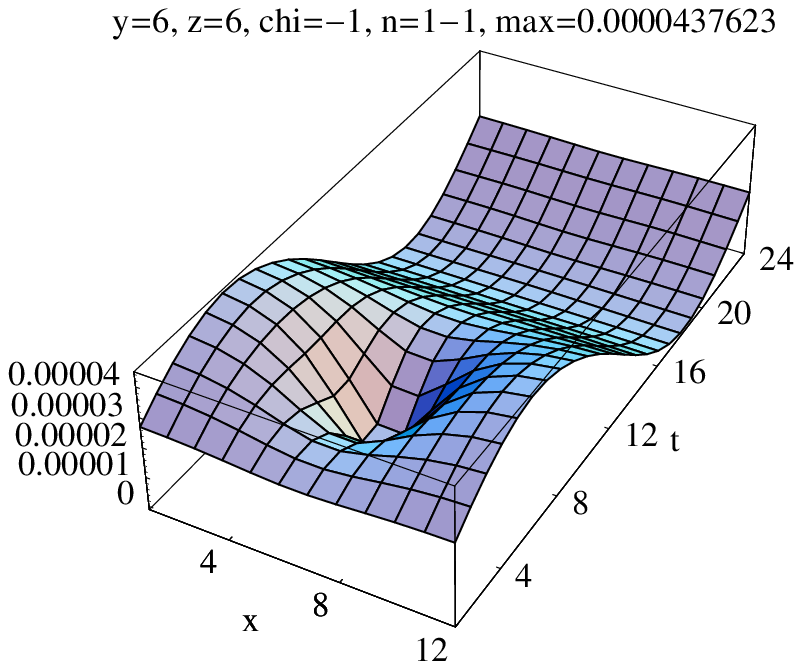}\\
		c)\includegraphics[width=.32\columnwidth]{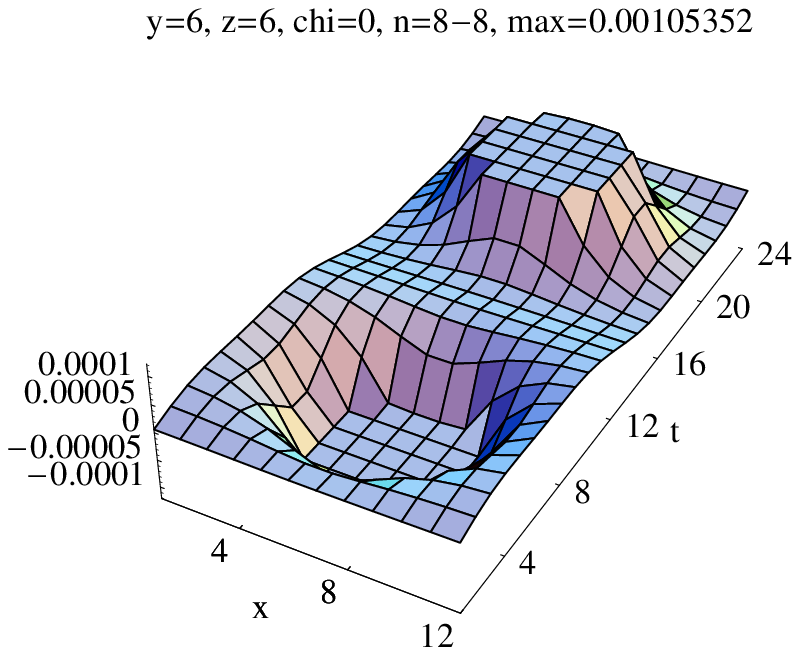}
		\includegraphics[width=.32\columnwidth]{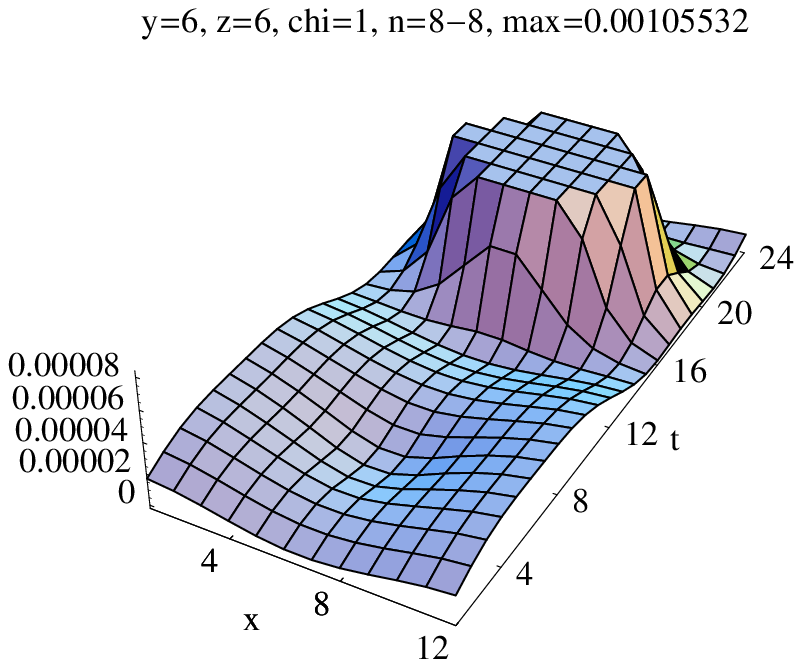}
		\includegraphics[width=.32\columnwidth]{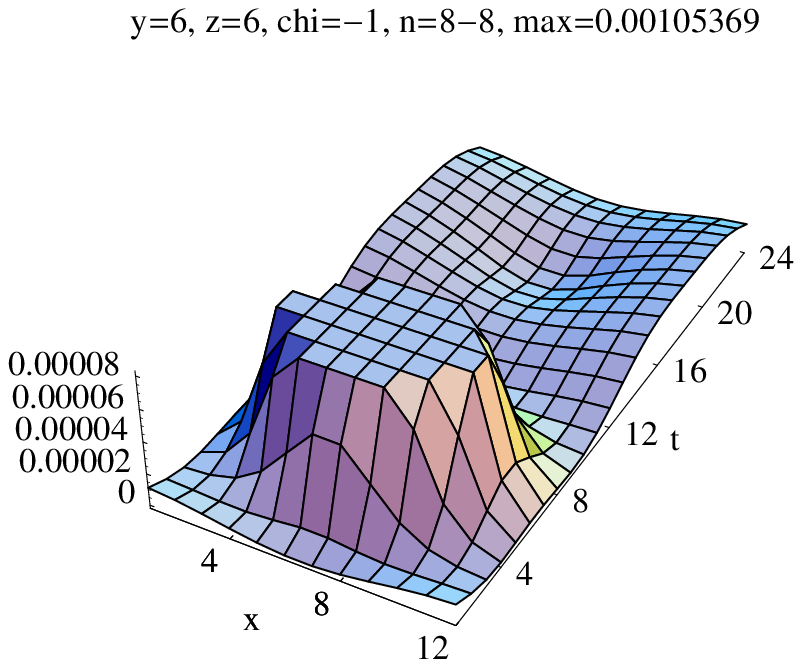}\\
		d)\includegraphics[width=.32\columnwidth]{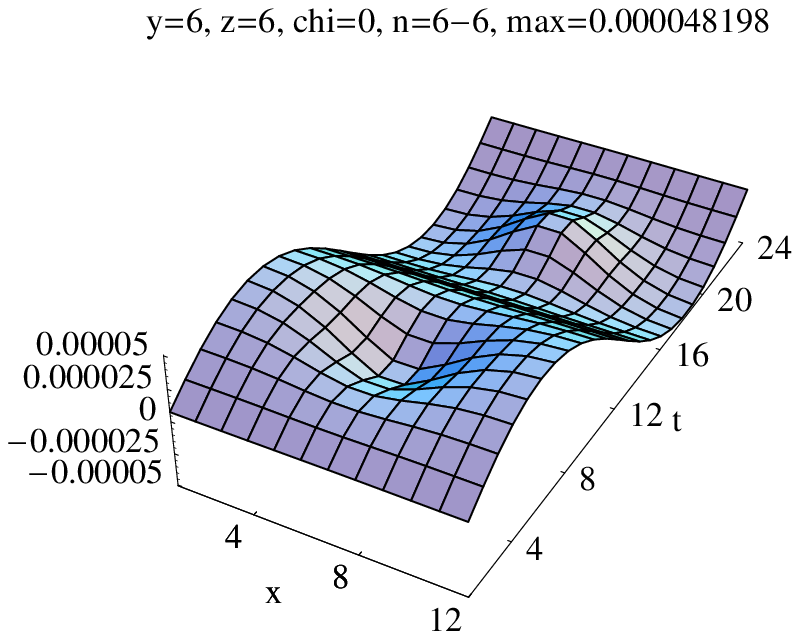}
		\includegraphics[width=.32\columnwidth]{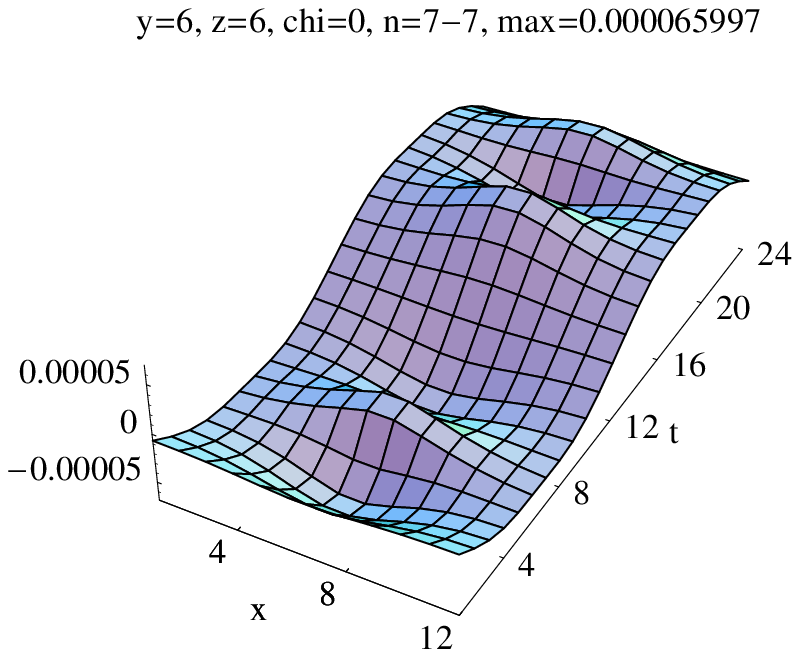}
		\includegraphics[width=.32\columnwidth]{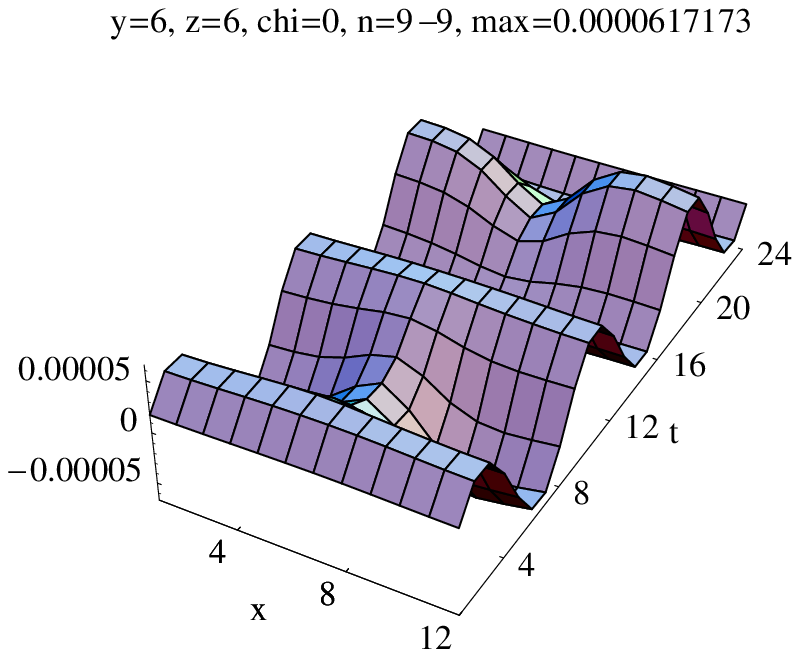}
	\caption{Chiral densities ($\rho_5$ left, $\rho_+$ center and
	$\rho_-$ right column) of the a) lowest (near-zero), b)
	second-lowest (nonzero) and c) eighth (nonzero) eigenmode of the
	overlap Dirac operator for an instanton--anti-instanton pair. d)
	$\rho_5$ of the sixth (left), seventh (center) and ninth (right)
	eigenmode.}
	\label{fig:iai}
\end{figure}

\begin{figure}
	\centering
		a)\includegraphics[width=.32\columnwidth]{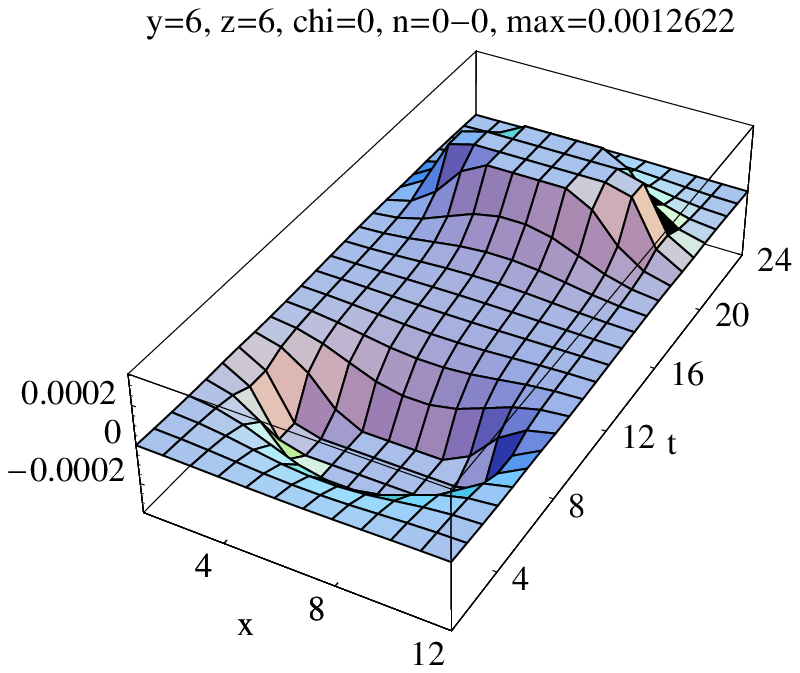}
		\includegraphics[width=.32\columnwidth]{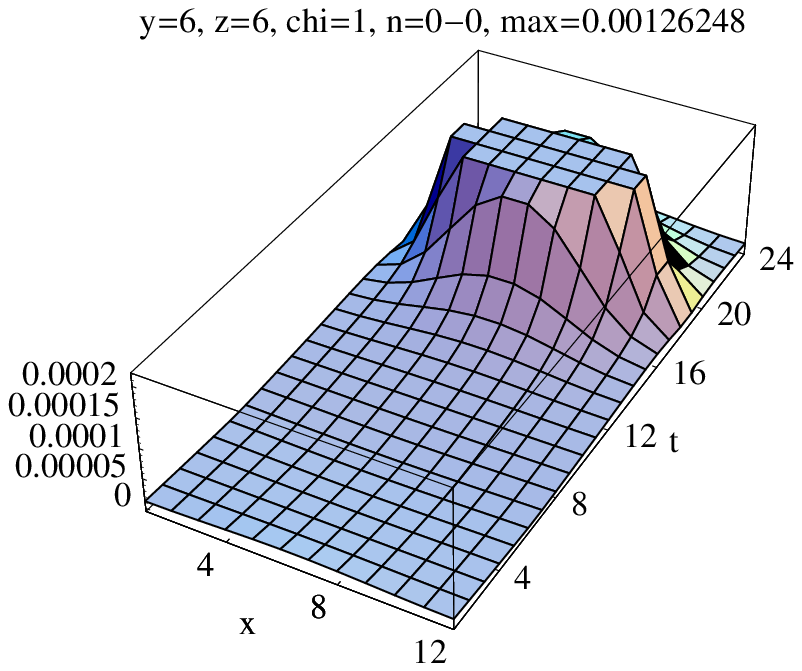}
		\includegraphics[width=.32\columnwidth]{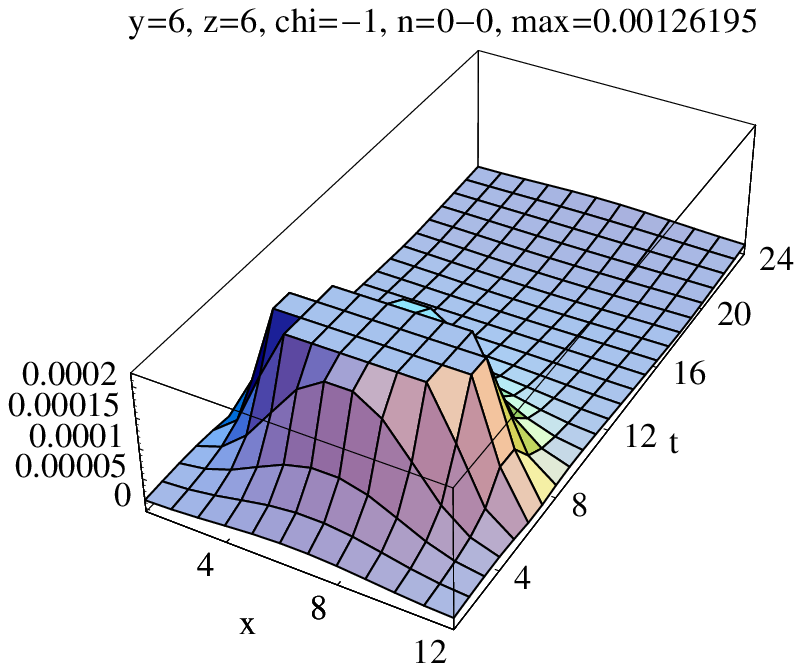}\\
	   b)\includegraphics[width=.32\columnwidth]{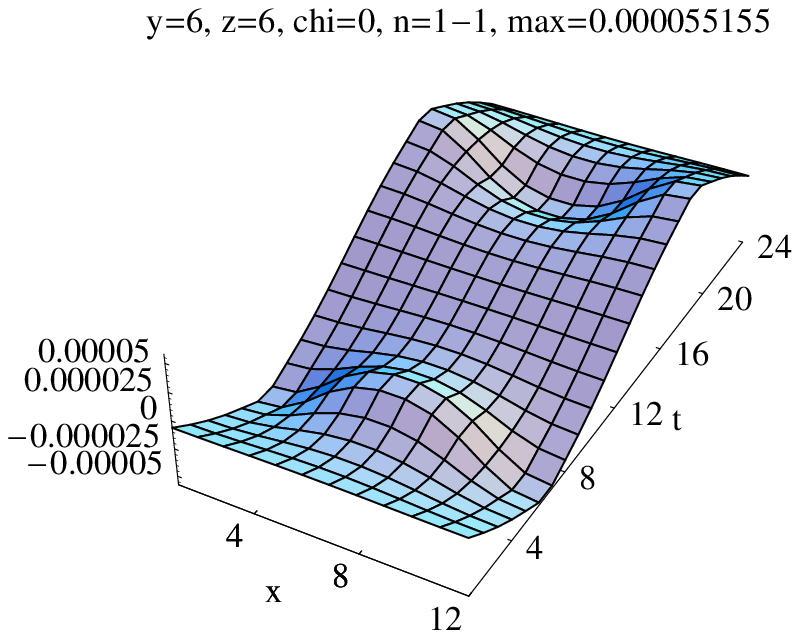}
		\includegraphics[width=.32\columnwidth]{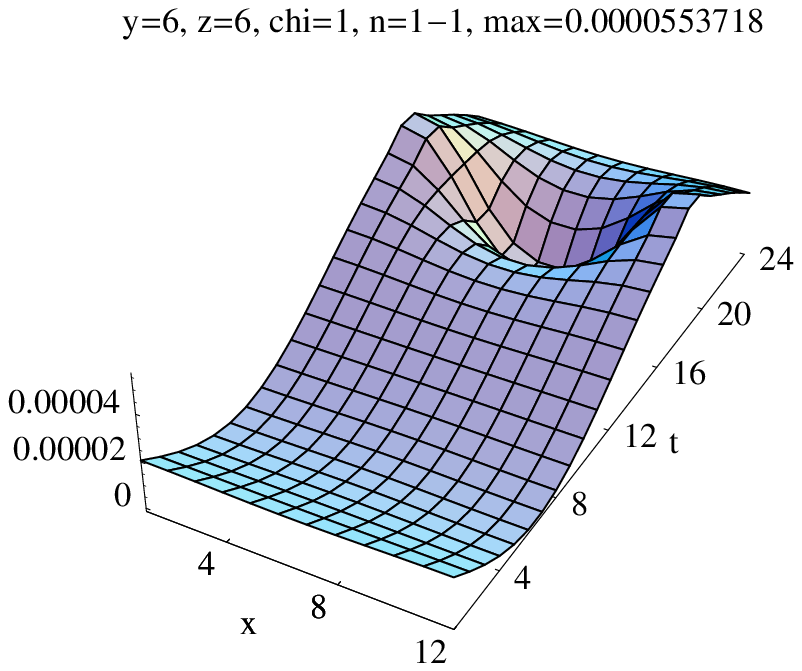}
		\includegraphics[width=.32\columnwidth]{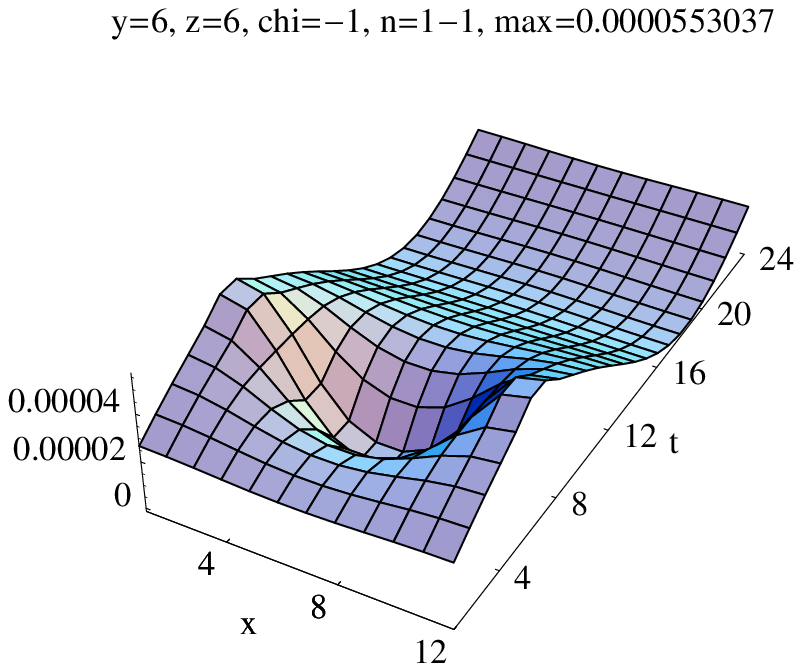}\\
		c)\includegraphics[width=.32\columnwidth]{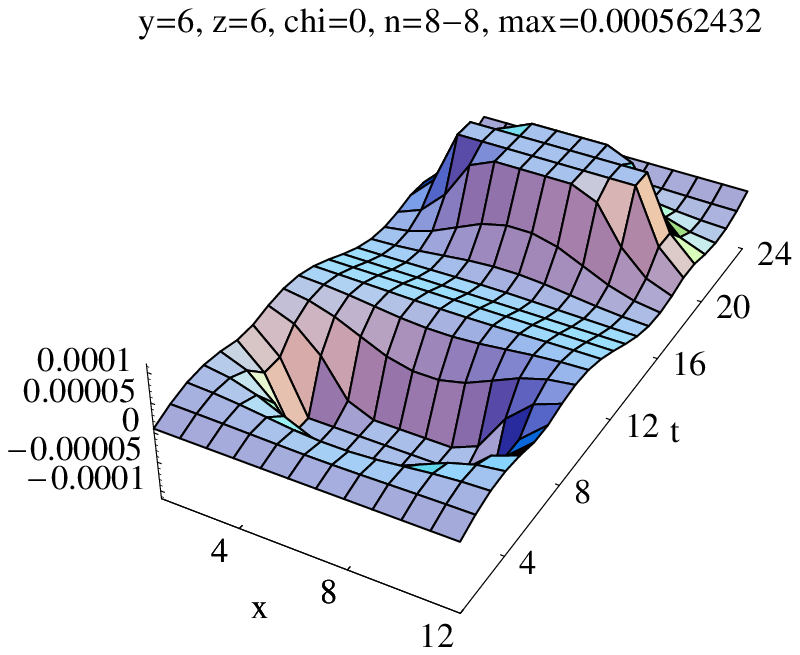}
		\includegraphics[width=.32\columnwidth]{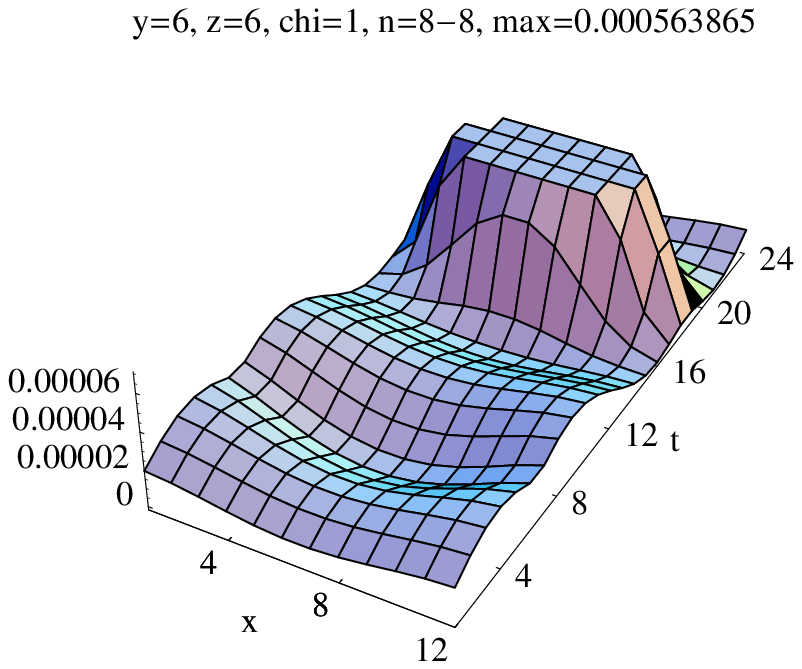}
		\includegraphics[width=.32\columnwidth]{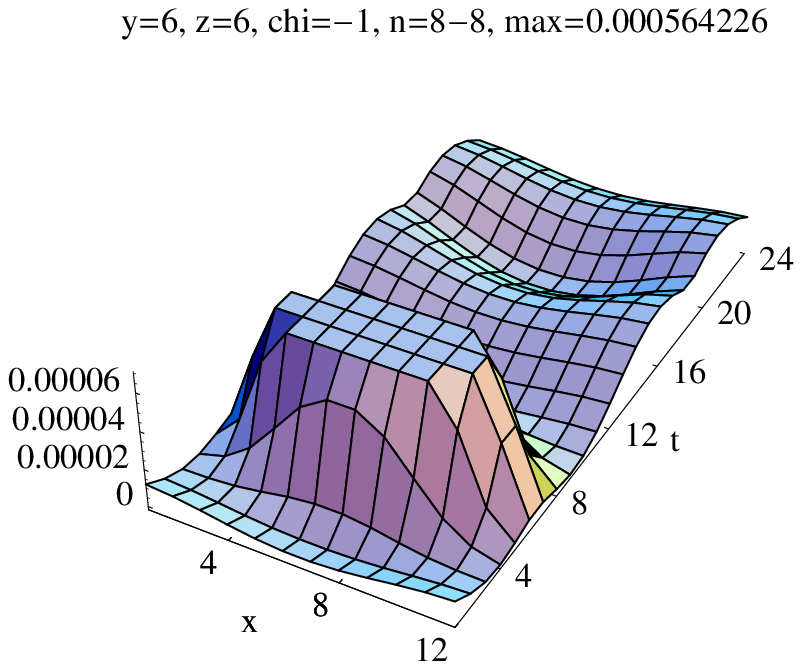}\\
	   d)\includegraphics[width=.32\columnwidth]{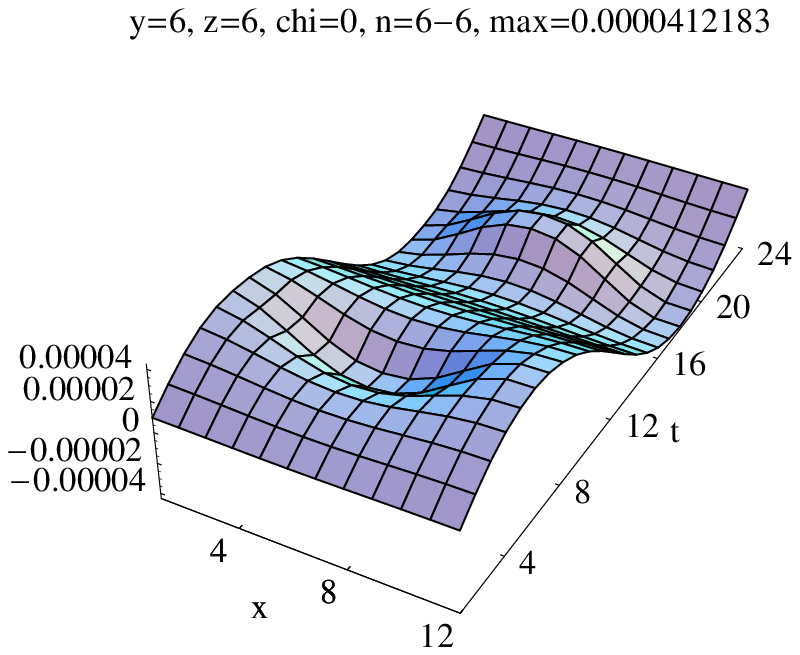}
		\includegraphics[width=.32\columnwidth]{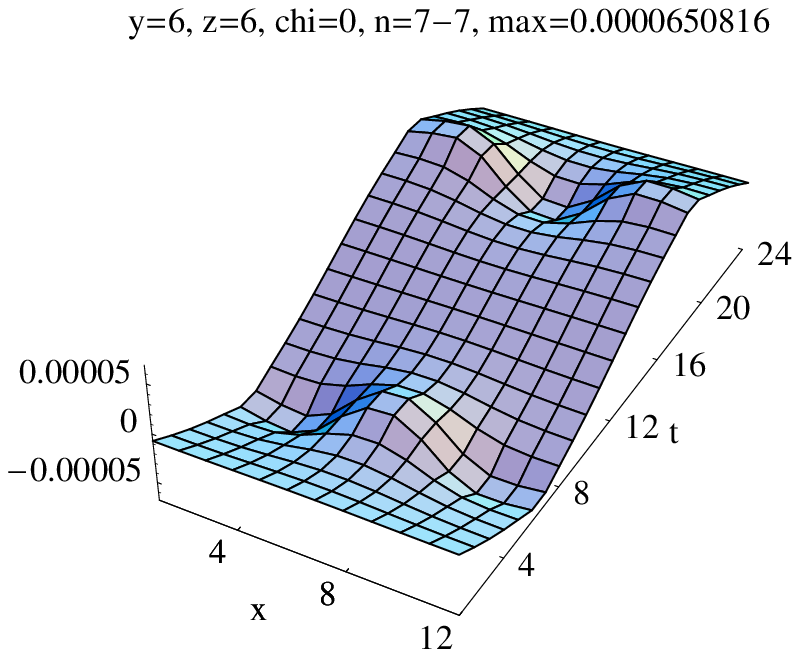}
		\includegraphics[width=.32\columnwidth]{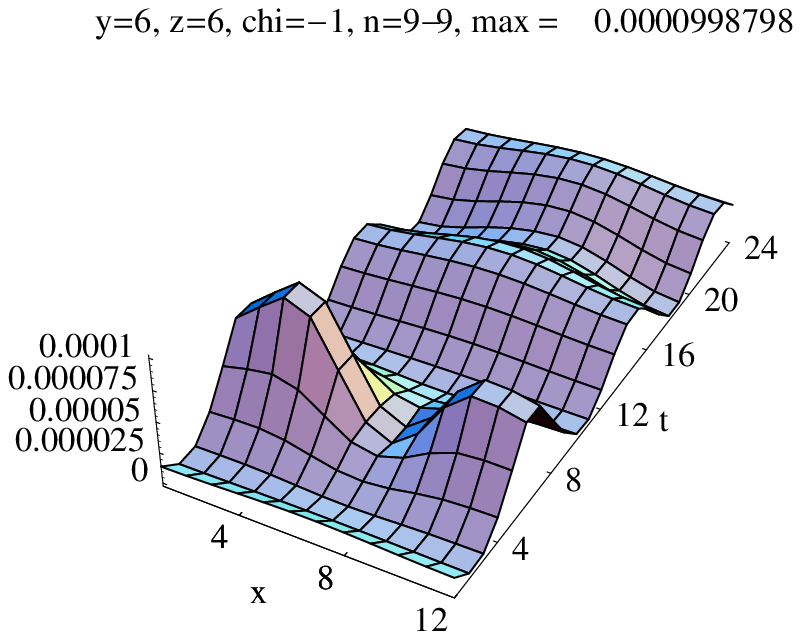}
	\caption{Same as Fig.~\protect\ref{fig:iai} but for a spherical
	vortex--anti-vortex pair.
	Chiral densities ($\rho_5$ left, $\rho_+$ center and
	$\rho_-$ right column) of the a) lowest (near-zero), b)
	second-lowest (nonzero) and c) eighth (nonzero) eigenmode.
	d) $\rho_5$ of the sixth (left), seventh (center) and ninth (right)
	eigenmode.}
	\label{fig:tat}
\end{figure}

The spherical vortex was introduced in~\cite{Jordan:2007ff} and analyzed in more 
detail in~\cite{Hollwieser:2010mj,Hollwieser:2012kb,Schweigler:2012ae}. It is 
constructed with $t$-links in a single time slice at fixed $t=t_i$, given by $U_t
(x^\nu) = \exp\left(\mathrm i\alpha(|\vec r-\vec r_0|)\vec r/r\cdot\vec\sigma
\right)$, where $\vec r$ is the spatial part of $x_\nu$. The profile function $
\alpha(r)$ changes from $\pi$ to $0$ in radial direction for the negative 
spherical vortex, or from $\pi$ to $2\pi$ for the positive (anti-)vortex. This 
gives a hedgehog-like configuration, since the color vector points in (or 
against) the radial direction at the vortex radius $R$. The hedgehog-like 
structure is crucial for our analysis. The $t$-links of the spherical vortex 
define a map $S^3 \to SU(2)$, characterized by a winding number $N=-1$ for 
positive (anti-) and $N=+1$ for negative spherical vortices. Obviously such 
windings influence the Atiyah-Singer index theorem giving a topological charge 
$Q=-1$ for positive and $Q=+1$ for negative spherical vortices (anti-vortices) 
and attract Dirac zero modes similar to instantons. In~\cite{Schweigler:2012ae} 
we showed that the spherical vortex is in fact a vacuum-to-vacuum transition in 
the time direction which can even be regularized to give the correct topological 
charge also from gluonic definitions. Fig.~\ref{fig:oevlsph}a shows that a 
spherical vortex has nearly exactly the same eigenvalues as an instanton. We 
further plot the spectra of instanton--anti-instanton, spherical vortex--anti-vortex and instanton--anti-vortex pairs. We again see nearly exactly the same 
eigenvalues for instanton or spherical vortex pairs, instead of two would-be zero 
modes there is a pair of near-zero modes for each pair. The chiral density plots in Fig.~
\ref{fig:iai} for the instanton--anti-instanton pair and Fig.~\ref{fig:tat} for 
the spherical vortex--anti-vortex pair show, besides the similar densities, that 
the near-zero modes are a result of two chiral parts corresponding to the two 
constituents of the pairs. The nonzero modes can be identified with the free 
overlap modes, as they show plane-wave behavior. In Fig.~\ref{fig:oevlsph}b we plot the 
eigenvalues of two (anti-)instantons and two spherical (anti-)vortices giving 
topological charge $Q=2$ ($Q=-2$) and therefore two zero modes, two vortex--anti-vortex pairs with two pairs of near-zero modes and a configuration with two vortices and an
anti-vortex ({\it i.e.}, a single vortex plus one vortex--anti-vortex pair)
giving one zero mode ($Q=1$) and one pair of near-zero modes. The results show 
that we may draw the same conclusions for spherical vortices as for instantons
concerning the creation of near-zero modes.


\section{Plane vortices}\label{sec:plane}

\begin{figure}[h]
	\centering
		a)\includegraphics[width=.48\columnwidth]{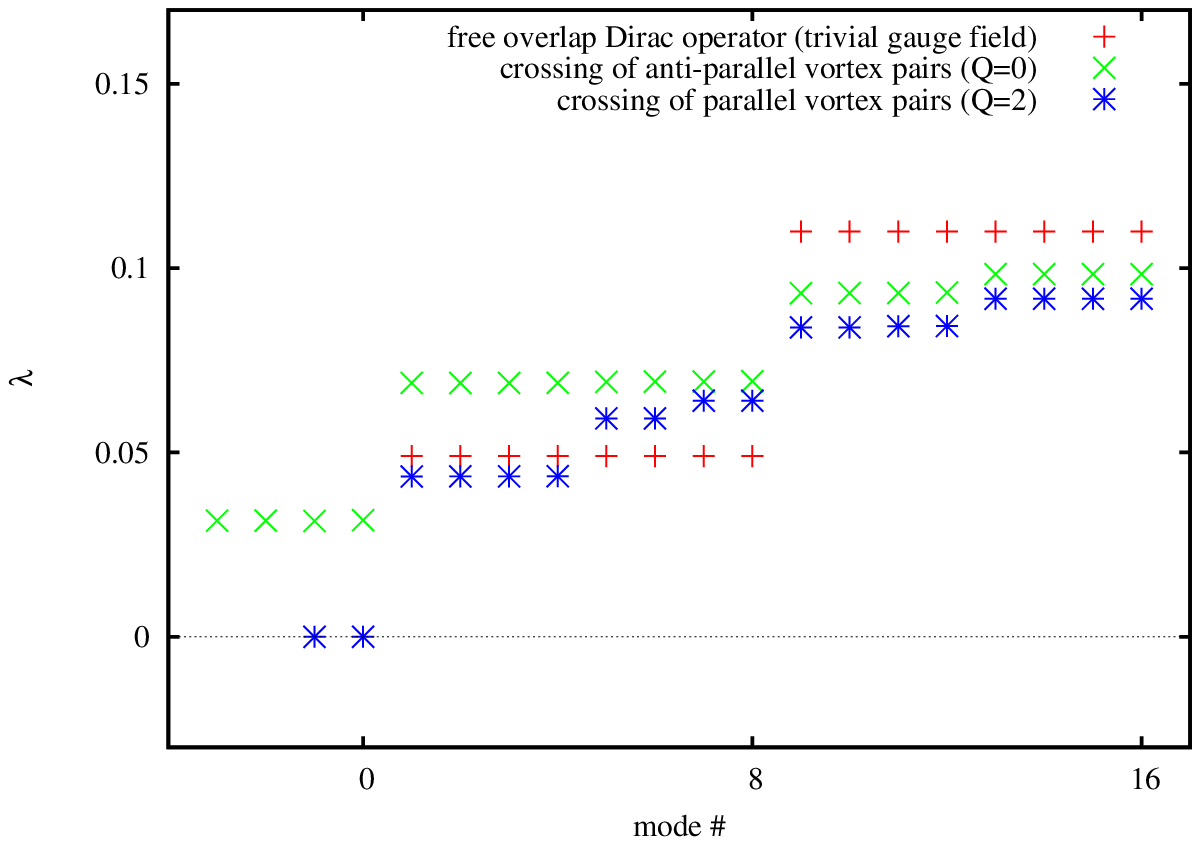}
		b)\includegraphics[width=.44\columnwidth]{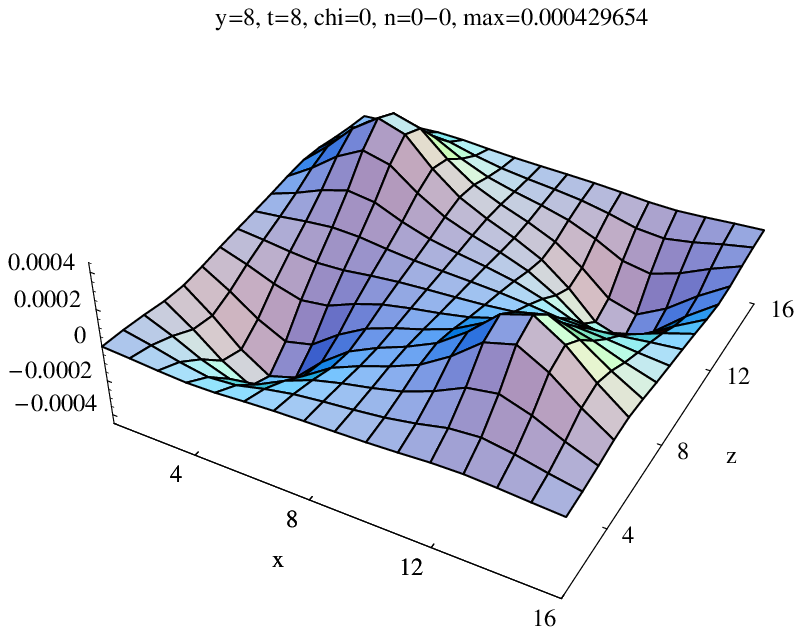}
	\caption{a) The lowest overlap eigenvalues for plane vortex
	configurations compared to the eigenvalues of the free (overlap)
	Dirac operator (red crosses) and spherical vortex configurations. 
	b) Chiral density in the
	intersection plane of all four near-zero modes of crossing flat vortex
	pairs with opposite flux direction ($Q=0$).}
	\label{fig:ovlart}
\end{figure}

We define plane vortices parallel to two of the coordinate axes by links varying 
in a $U(1)$ subgroup of $SU(2)$. This $U(1)$ subgroup is generated by one of the 
Pauli matrices $\sigma_i$, {\it i.e.}, $U_\mu=\exp(\mathrm{i} \phi \sigma_i)$.  
Upon traversing a vortex sheet, the angle $\phi$ increases or decreases by
$\pi$ within a finite thickness of the vortex. Since we use periodic (untwisted) 
boundary conditions for the links, vortices occur in pairs of parallel sheets, 
each of which is closed by virtue of the lattice periodicity. We call vortex
pairs with the same vortex orientation parallel vortices and vortex pairs of
opposite flux direction anti-parallel. If thick, planar vortices intersect 
orthogonally, each intersection carries a topological charge $|Q|=1/2$, whose 
sign depends on the relative orientation of the vortex fluxes~\cite
{Engelhardt:1999xw}. For intersecting parallel vortices we get two real zero 
modes, according to the total topological charge $Q=2$ of the four intersections. 
These modes we analyzed in~\cite{Hollwieser:2011uj}, they peak at least at two of 
the four topological charge contributions of $Q=1/2$. If we intersect anti-parallel vortex pairs orthogonally we get two intersection points with 
topological charge $Q=+1/2$ and two intersection points with topological charge 
$Q=-1/2$, hence total $Q=0$. For such a configuration we get four real near-zero 
modes, with local chirality peaks at the intersection points, according to their 
topological charge contribution, see Fig.~\ref{fig:ovlart}b. The mechanism of 
Sec.~\ref{sec:spher} or the analog instanton liquid model does not directly apply 
to the case of planar vortices, since there are no localized lumps of topological 
charge $Q=\pm 1$. 
But vortex intersections with topological charge $Q=\pm 1/2$ might be
related to merons and calorons. In any case, we showed that they, too, can
create a finite density of near-zero modes and thus cause $\chi$SB via
the Banks-Casher relation.

\vspace{-5mm} 

\section{Conclusions}
Fermions do not seem to make much of a difference between instantons and
spherical vortices and the instanton liquid model can be extended to colorful 
spherical center vortices. Further also vortex intersections attract (would-be) 
zero modes which contribute via interactions to a finite density of near-zero 
modes with local chiral properties, {\it i.e.}, local chirality peaks at 
corresponding topological charge contributions. In Monte Carlo configurations we 
do not, of course, find perfectly flat or spherical vortices, as one does not find
perfect instantons. The general picture of topological charge from vortex
intersections, writhing points and even color structure contributions can provide a general picture of $\chi$SB: just like instantons any source of
topological charge can attract (would-be) zero modes and produce a finite
density of near-zero modes leading to chiral symmetry breaking via the
Banks-Casher relation. Here one also has to ask what could be the dynamical
explanation of $\chi$SB. We can try the conjecture that only a combination
of color electric and magnetic fields leads to $\chi$SB, electric fields
accelerating color charges and magnetic fields trying permanently to
reverse the momentum directions on spiral shaped paths. Such reversals of
momentum keeping the spin of the particles should especially happen for
very slowly moving color charges. Alternatively we could argue that
magnetic color charges are able to flip the spin of slow quarks, {\it i.e.}
when they interact long enough with the vortex structures. Finally, it seems that 
vortices not only confine quarks into bound states but also break chiral symmetry.
For more details see~\cite{Hollwieser:2013xja}. 

%\acknowledgments{We thank {\v S}tefan Olejn\'{\i}k and Michael Engelhardt for helpful discussions. This research was supported by the Austrian Science Fund FWF (``Fonds zur F\"orderung der wissenschaftlichen Forschung'') under Contract No. P22270-N16 (R.H.).}

\bibliographystyle{utphys}
\bibliography{../literatur}
%\bibliography{literatur}
%\bibliography{scsb}

\end{document}